\documentclass{article}

 \usepackage{amsmath}
 \usepackage{bm}
 \usepackage{graphicx}
\usepackage{authblk}

\begin{document}

\title{The inadequacy of a magnetohydrodynamic approach  to the Biermann Battery}

  \author[$1$]{C. P. Ridgers}
  \author[$1$]{C. Arran}
  \author[$2$]{J. J. Bissell}
  \author[$3$]{R. J. Kingham}

\affil[$1$]{York Plasma Institute, Department of Physics, University of York, Heslington, York, North Yorkshire, YO10 5DD, United Kingdom}
\affil[$2$]{Department of Electronic Engineering, University of York, Heslington, York, North Yorkshire, YO10 5DD, United Kingdom}
\affil[$3$]{Blackett Laboratory, Imperial College London, Prince Consort Road, South Kensington, London, SW7 2AZ, United Kingdom}

\maketitle

\begin{abstract}
  Magnetic fields can be generated in plasmas by the Biermann battery when the electric field produced by the electron pressure gradient has a curl.  The commonly employed magnetohydrodynamic (MHD) model of the Biermann battery breaks down when the electron distribution function is distorted away from Maxwellian.  Using both MHD and kinetic simulations of a laser-plasma interaction relevant to inertial confinement fusion we have shown that this distortion can reduce the Biermann-producing electric field by around 50\%. More importantly, the use of a flux limiter in an MHD treatment to deal with the effect of the non-Maxwellian electron distribution on electron thermal transport leads to a completely unphysical prediction of the Biermann-producing electric field and so results in erroneous predictions for the generated magnetic field. 
\end{abstract}


\section{Introduction}

Magnetic fields are ubiquitous in plasmas. The Biermann battery is one important mechanism by which these magnetic fields are generated \cite{Biermann_50}.  A simplistic explanation for this mechanism is that it arises from the electric field caused by an electron pressure gradient.  If this electric field has a curl, then Faraday's law predicts the generation of a B-field.  The Biermann battery has been proposed as a mechanism for magnetic field generation in a wide range of plasma environments from astrophysical shocks \cite{Gregori_04} to laser-driven inertial confinement fusion (ICF) experiments \cite{Glenzer_99,Farmer_17}.  The simple interpretation of the Biermann battery being caused by the curl of the electric field due to the electron pressure gradient only holds for plasmas where the electrons are in local thermodynamic equilibrium (LTE).  One requirement for this is that the mean free path of the electrons should be much smaller than the scale-length of variation of the electron pressure gradient.  This assumption often breaks down and the electron transport becomes nonlocal depending on conditions in distant regions of the plasma \cite{Bell_81}.  In this case the electrons are not in LTE, their distribution function is non-Maxwellian and a kinetic description of the Biermann battery is required.  If the distortion of the distribution away from Maxwellian due to nonlocality is taken into account it has been shown that the B-field generated by the Biermann battery is reduced \cite{Kingham_02,Kingham_04_2,Sherlock_20}. 

Here we will use kinetic simulations to show that the commonly employed LTE formulation of the Biermann battery (as the curl of the electron-pressure gradient electric field) can be erroneous due to an inaccurate treatment of nonlocality.  In particular we will focus on conditions realisable experimentally with high intensity lasers and of importance to ICF.  We will show that while the direct nonlocal correction to the Biermann-producing electric field\footnote{As a matter of terminology we resist referring to this as the field due to the electron pressure gradient from now on as this identification is only true when the electron distribution function is Maxwellian.} from the distortion of the electron distribution is significant, more important are errors introduced by an LTE treatment of the electron thermal transport predicting temperature gradients which are far too steep. Non-LTE electron thermal transport is a well known problem in the context of laser-produced plasmas \cite{Bell_81} and ICF \cite{Rosen_11} and has been more recently considered in the context of magnetic confinement fusion \cite{Omotani_13,Chankin_19,Wigram_20,Mijin_20}. Its effect on the Biermann battery, however, has not been considered and will be elucidated here. 

This paper will be organised as follows.  In section \ref{Kinetic_Biermann} we will discuss the specific nonlocal effects on the Biermann battery.  Section \ref{simulations} will show a comparison of kinetic and magnetohydrodynamic (MHD) simulations, the latter assuming LTE, demonstrating the importance of nonlocal effects on the Biermann battery in laser-plasma interactions relevant to ICF. We will suggest how these effects may be observed experimentally.

\section{Nonlocal Effects on the Biermann Battery}
\label{Kinetic_Biermann}

\subsection{The direct effect of distortion of the electron distribution function}

A kinetic description of the Biermann battery has been given previously \cite{Kingham_02} but will be reiterated here for completeness.  We start from the equation describing the time evolution of the electron distribution function $f(\mathbf{x},\mathbf{v},t)$ in phase space $(\mathbf{x},\mathbf{v})$, initially in the absence of a magnetic field $\mathbf{B}$.

\begin{equation}
\frac{\partial f}{\partial t} + \mathbf{v}\cdot\nabla_{\mathbf{x}}f- \frac{e\mathbf{E}}{m_e} \cdot\nabla_{\mathbf{v}}f = \hat{C}(f) + \hat{H}(f). \nonumber
\end{equation}

\noindent $\nabla_{\mathbf{x}}$ \& $\nabla_{\mathbf{v}}$ are gradients in $\mathbf{x}$ \& $\mathbf{v}$ space respectively.   $\mathbf{E}$ is the electric field resulting from collective plasma processes (including the Biermann-producing electric field but not limited to it).  $\hat{C}$ is the collision operator, we use the Fokker-Planck operator and this combination is referred to as the Vlasov-Fokker-Planck (VFP) equation. In addition we include the operator $\hat{H}$ describing heating of the plasma by an external source, for example a laser.  We assume that while the electron distribution function is far from Maxwellian, it is not very anisotropic, i.e. we are in a weakly collisional regime where the mean free path of the electrons is not much longer than the pressure scale-length. In this case we may expand $f$ in Cartesian tensors, keeping only the isotropic part $f_0$ and first order anisotropy $\mathbf{f}_1$, $f(\mathbf{x},\mathbf{v},t) = f_0(\mathbf{x},v,t)+\mathbf{f}_1(\mathbf{x},v,t)\cdot\mathbf{v}/v$ \cite{Johnston_60}.  Substituting this into the Vlasov-Fokker-Planck equation yields

\begin{eqnarray}
\label{f0_eqn}
\frac{\partial f_0}{\partial t} + \frac{v}{3}\nabla_{\mathbf{x}}\cdot\mathbf{f}_1 - \frac{e}{3m_ev^2}\frac{\partial}{\partial v}(v^2\mathbf{E}\cdot\mathbf{f}_1) = \hat{C}_{ee}(f_0) + \hat{H}(f_0) \\
\label{f1_eqn}
\frac{\partial \mathbf{f}_1}{\partial{t}} +v\nabla_{\mathbf{x}}f_0 - \frac{e\mathbf{E}}{m_e}\frac{\partial f_0}{\partial v} = -\nu_{ei}\mathbf{f}_1.
\end{eqnarray}

\noindent We have only included the effect of electron-electron collisions on $f_0$ as these dominate electron energy exchange and only electron-ion collisions on $\mathbf{f}_1$ as for high $Z$ these dominate electron angular scattering; $\nu_{ei}=YZn_e\mbox{ln}\Lambda_{ei}/v^3$ is the electron-ion collision frequency ($n_e$ is the electron number density, $Z$ the ionic charge, $\mbox{ln}\Lambda_{ei}$ the Coulomb logarithm for electron-ion collisions and $Y=4\pi[e^2/4\pi\epsilon_0m_e]^2$).  While equations (\ref{f0_eqn}) and (\ref{f1_eqn}) include terms beyond the Biermann-producing electric field, they are stated here as they are the equations solved by the kinetic code IMPACT \cite{Kingham_04} used to perform the simulations in section \ref{simulations} below\footnote{This code also includes the affect of magnetic fields but as none are present in our simulations they are omitted here for simplicity.}.

To derive a kinetic expression for the Biermann-producing electric field we consider equation (\ref{f1_eqn}), neglecting electron inertia (i.e. $\partial\mathbf{f}_1/\partial t$).  We then multiply this equation by $v^6$ and integrate.  In this case the term on the right-hand side (i.e. containing the collision frequency $\nu_{ei}$) is proportional to the current, if we assume this is zero we arrive at

\begin{equation}
\label{nonlocal_E}
\mathbf{E} = -\frac{m_e}{6e}\frac{\nabla_{\mathbf{x}}(n_e\langle v^5\rangle)}{n_e \langle v^3 \rangle}.
\end{equation}

\noindent Here the velocity averages of the distribution function are defined as $\langle v^n\rangle = (4\pi/n_e)\int_0^{\infty} v^{n+2}f_0 dv$.

Equation (\ref{nonlocal_E}) holds for non-Maxwellian $f_0$.  We can recover the standard result for a plasma in LTE by substituting $f_0=f_M$, where $f_M$ is Maxwellian. $f_M = [n_e/(\pi^{3/2}v_T^3)]\exp(-v^2/v_T^2)$ and $v_T$ is the electron thermal speed, related to the electron temperature by $v_T = (2T_e/m_e)^{1/2}$.  It can be shown that $\langle v^n \rangle_M = (4\pi/n_e)\int_0^{\infty} v^{n+2}f_M dv  = (2\sqrt{\pi})\Gamma[(n+3)/2]v_T^{n/2}$ and so

\begin{equation}
  \label{local_E}
\mathbf{E}_M = -\frac{1}{e}\left(\frac{\nabla_{\mathbf{x}}P_e}{n_e} + \frac{3}{2}\nabla_{\mathbf{x}}T_e \right),
\end{equation}

\noindent where $\mathbf{E}_M$ is the electric field assuming the electron distribution function is Maxwellian. The first, electron pressure gradient term gives rise to the Biermann battery while the second, thermoelectric term is curl-free and so usually neglected (though it will be retained in our MHD simulations).

\subsection{The indirect effect of electron thermal transport}
\label{Indirect_kinetic _Biermann}

Equations (\ref{nonlocal_E}) and (\ref{local_E}) demonstrate that a non-Maxwellian electron distribution function has a direct effect on the Biermann-producing electric field.  Another, indirect effect is that magnetohydrodynamic and kinetic approaches will predict different macroscopic plasma conditions such as electron temperature, which will then have a significant impact on their prediction for the Biermann-producing electric field. We use the general definition of temperature as the second moment of the electron distribution $T_e = \langle m_ev^2/2\rangle$ and thus it is defined even when the distribution is non-Maxwellian.

The reason for the large discrepancy between the electron temperature in MHD and kinetic simulations is due to kinetic non-local effects on the electron heat flow $\mathbf{q}_e$, the same as for the Biermann-producing electric field (in fact more important as the heat flow depends on faster, less collisional electrons).  The most commonly applied fix to this problem in MHD is to artificially limit the electron heat flow when the model predictions become unphysically large (this does not occur in the kinetic model which accurately captures non-local effects), typically limiting the heat flow to some fraction of the free-streaming limit $q_{fs}$ \cite{Malone_75}.  The free-streaming limit expresses the maximum heat flow which could be obtained if all electrons were flowing collisionlessly down the temperature gradient at the thermal speed, usually\footnote{Alternative pre-factors to $1/2$ are also used, $q_{fs}=n_em_ev_T^3/3$ and $n_em_ev_T^3$are also common.} expressed as $q_{fs}=n_em_ev_T^3/2$.  Such a flux limiter $f$ may be implemented in a variety of ways but a harmonic average is usual such that $1/|\mathbf{q}_e|=1/|\mathbf{q}_{e,\mbox{MHD}}|+1/(fq_{fs})$ where $\mathbf{q_{e,\mbox{MHD}}}$ is the electron heat flow predicted by MHD.

In practice it is not possible to fully capture nonlocal effects on thermal transport with a flux limiter.  The degree of nonlocality varies in space and time and, more importantly for the Biermann battery,  the flux limiter has the effect of artificially steepening the electron temperature profile and so pressure profile dramatically. In reality hot, long mean free path electrons would stream ahead of the heat front and smooth these gradients but this pre-heat phenomenon is not captured at all by a flux limiter.  In section \ref{simulations}\ref{sim_results} we will see that this artifical steepening causes MHD to over-predict the Biermann battery.

\section{Comparison of Kinetic to MHD simulations}
\label{simulations}

\subsection{Simulation setup}

We will compare VFP to MHD simulations of a situation realisable experimentally with high intensity lasers. We will consider a laser heating an underdense gas, a situation of direct applicability to the gas fill in a hohlraum in the indirect drive ICF scheme \cite{Gregori_04}. For relevance to the gas-fill the gas density was chosen to be $1.5\times 10^{19}$ cm$^{-3}$.  The gas was composed of fully ionised nitrogen and the Coulomb logarithm $\ln\Lambda_{ei}$ was set to eight.  It was assumed to be at a temperature of 20 eV before being heated by the laser.  The laser was chosen to have peak intensity $2\times 10^{14}$ Wcm$^{-2}$, a Gaussian spatial profile with FWHM 33.3 microns and a flat-top 1 ns temporal profile.  The simulations were performed in one spatial dimension in planar geometry.

We use the VFP code IMPACT \cite{Kingham_04}.  IMPACT solves equations (\ref{f0_eqn}) and (\ref{f1_eqn})\footnote{With the addition of small correction terms due to bulk ion motion not given here for simplicity, see Refs. \cite{Epperlein_88} \& \cite{Ridgers_08}} for the time evolution of both the isotropic part $f_0$ and first-order anisotropy $\mathbf{f}_1$ of the electron distribution function.  IMPACT also directly solves for the electric field and, as it allows the electron distribution to be non-Maxwellian, naturally includes non-local effects on the Biermann-producing electric field.  Our IMPACT simulations include the heating operator $\hat{H}$ modelling inverse bremsstrahling heating \cite{Langdon_80}.

Two comparisons to IMPACT simulations will be made to determine kinetic effects on the Biermann-producing electric field identified in section \ref{Kinetic_Biermann}.  Firstly, the direct effect of the distortion of the distribution away from Maxwellian can be determined by taking the electron temperature (defined as $\langle m_ev^2/2\rangle$) and number density ($\langle v^0 \rangle$) profiles from the IMPACT simulation and applying equation (\ref{local_E}) to determine the electric field we would predict if the distribution function were locally Maxwellian for these plasma conditions.  To determine the indirect effect that the plasma conditions evolve differently in a kinetic and MHD model due to kinetic effects on the thermal transport we can compare IMPACT simulations to those using the MHD code CTC \cite{Bissell_12}.  CTC solves the equations for conservation of mass, momentum and energy in the plasma along with the local Ohm's law.  An inverse bremsstrahlung heating operator is included in the energy equation to model laser heating.  CTC applies a flux limiter to the heat flow as described previously.

In the CTC simulations the spatial grid was 800 microns in size represented by 240 grid cells with the laser spot at the centre (as CTC only has periodic boundary conditions).  1 ns of laser heating was simulated using 150,000 time steps (giving a time step of 66.7 fs).  The IMPACT simulations used a spatial grid 878 microns in size, represented by 200 grid cells.  The laser spot was centred at one boundary and reflective boundary conditions were applied.  The $v$-grid extended to 20$v_{T0}$, where $v_{T0}$ is the thermal speed for the initially unheated plasma (at 20 eV).  The timestep size was set to 27.6 fs (the electron-ion collision time for thermal electrons in the unheated plasma).  \footnote{Note that we compared inverse bremsstrahlung heating of a plasma in 0D in IMPACT and CTC simulations to ensure they were being heated at the same rate when the different transport models did not affect the result.}

\subsection{Simulation results}
\label{sim_results}

\begin{figure}
\centering\includegraphics[width=5in]{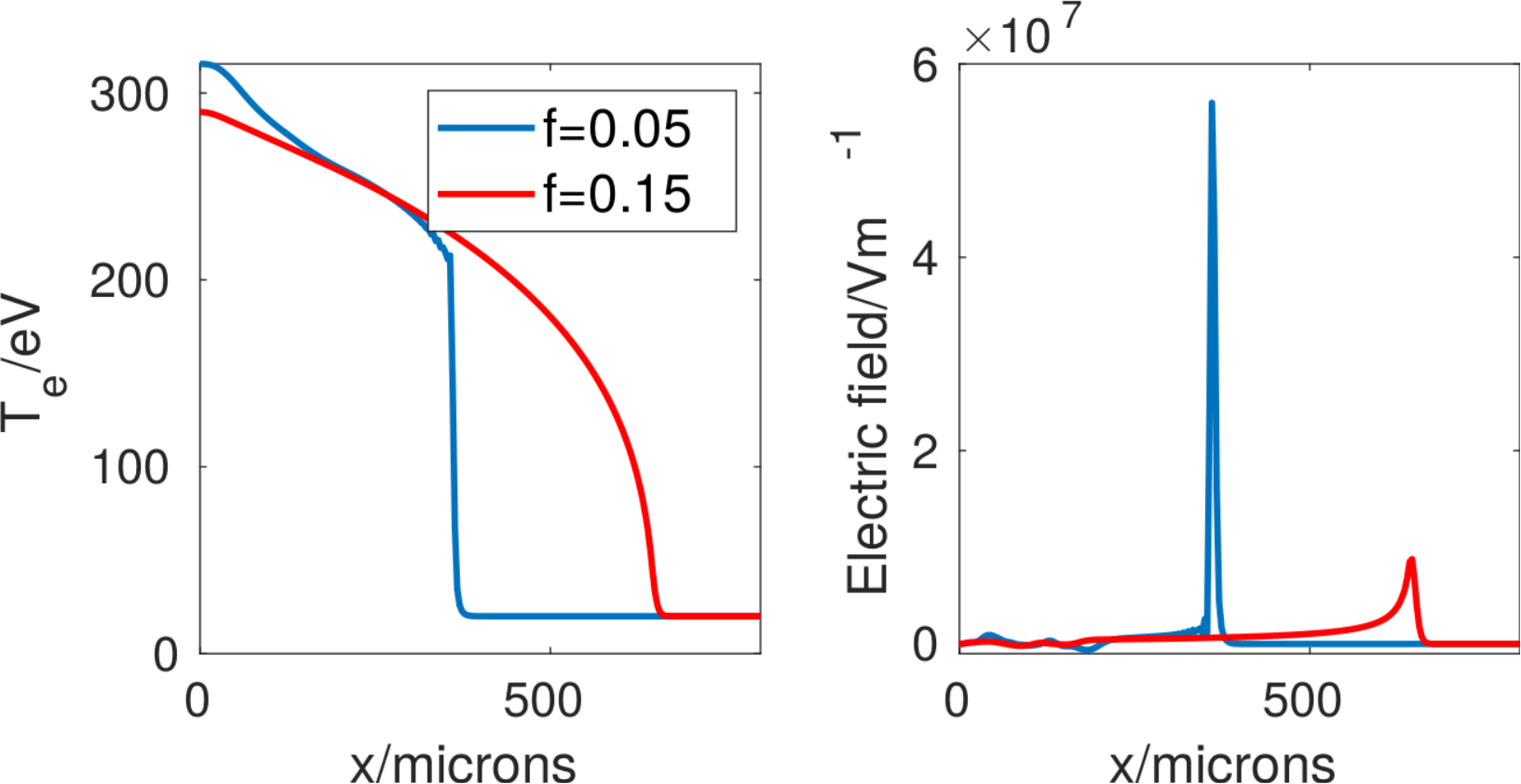}
\caption{Left: electron temperature profile from CTC after 750 ps laser heating using different flux limiters, $f=0.05$ (blue line) \& $f=0.15$ (red lines). Right: corresponding electric field (in the $x$-direction) for $f=0.05$ (blue line) \& $f=0.15$ (red line).}
\label{temp_efield_CTC}
\end{figure}

Figure \ref{temp_efield_CTC} shows the electron temperature and electric field, produced by CTC, after 1 ns of laser heating, using commonly employed values for the thermal flux limiter $f=0.15$ and $f=0.05$ \cite{Rosen_11,Jones_17}.  From this figure it is clear that the thermal flux limiter makes a large difference to the electric field.  Particularly in the $f=0.05$ case the flux limiter causes substantial steepening of $T_e$.  Limiting the heat flux creates an artificial transport barrier and thus the steep fall in $T_e$ clearly shown in figure \ref{temp_efield_CTC}.  In the case of CTC simulations, the Biermann-producing electric field is determined by $n_e$ and $T_e$ using equation (\ref{local_E}). This is dominated by the artificially steep part of the $T_e$ profile resulting in an erroneously large, spatially localised electric field. We can see this directly from equation (\ref{local_E}) which, on utilising the equation of state $P_e=n_eT_e$, becomes $\mathbf{E}\approx-(5/2e)\nabla_{\mathbf{x}}T_e$ as the electron density does not vary substantially. This is reasonable as while the density cavitates by 30\% from the laser heated region after 1 ns in the $f=0.05$ case (which shows most density cavitation) the variation over the region containing the flux limiter induced transport barrier is only 13\%. The CTC results plotted in figure \ref{temp_efield_CTC} show that the temperature drops by 150 eV over 6.8 microns (two spatial grid cells) due to the transport barrier introduced by the flux limiter.  We therefore predict $|\mathbf{E}|\sim 6\times 10^7$ Vm$^{-1}$, in agreement with the CTC simulation results for the electric field and verifying that the flux limiter causes this therefore spurious E-field.  The steepening is less pronounced in the $f=0.15$ case, however the electric field is still dominated by the artificially steepened part of the electron temperature profile.  For $f=0.15$, after 1 ns of heating CTC predicts a drop in $T_e$ of 50 eV over 17 microns at the artificial transport barrier, giving an estimate from equation (\ref{local_E}) of $|\mathbf{E}| \sim 7\times 10^6$ Vm$^{-1}$ again in agreement with simulations 

\begin{figure}
\centering\includegraphics[width=5in]{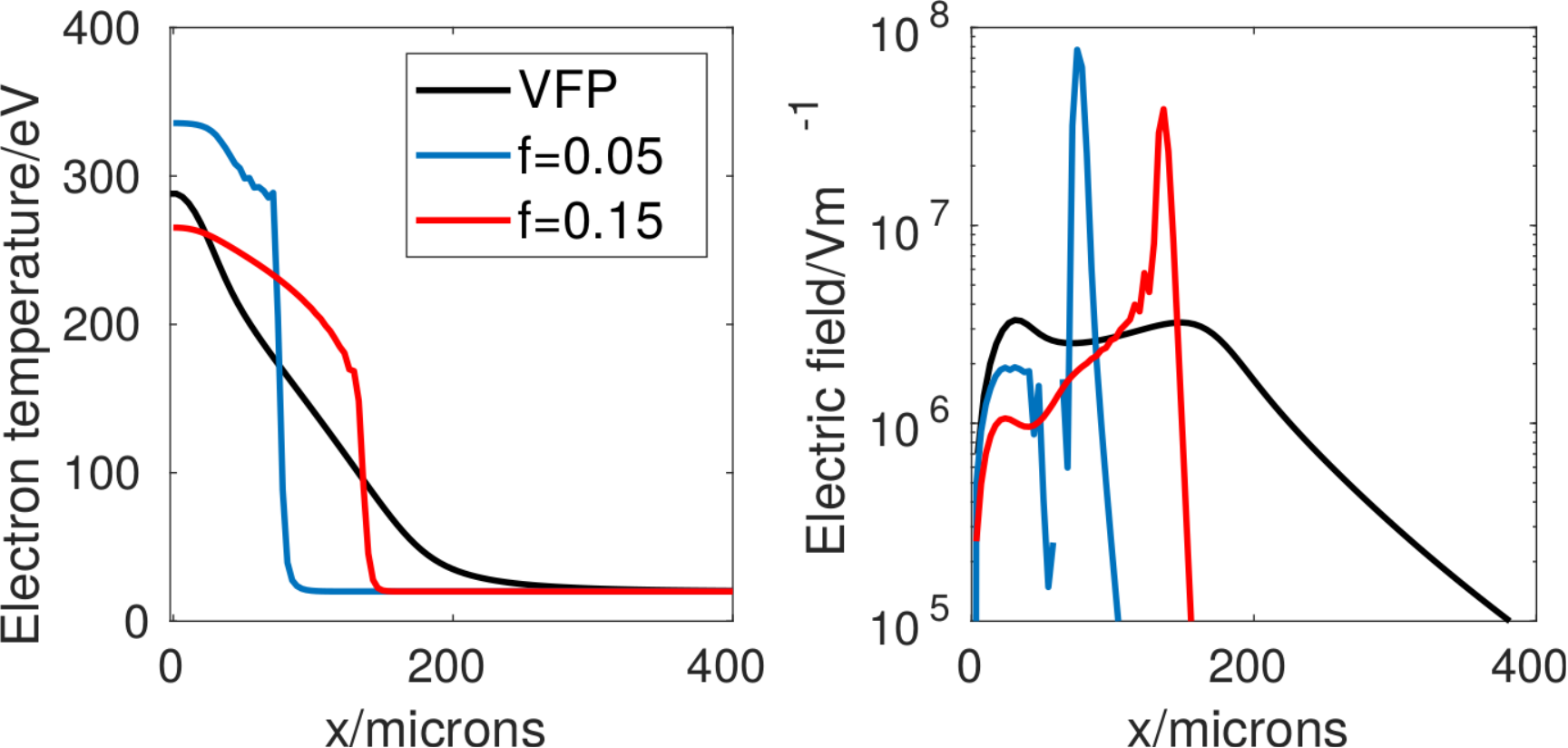}
\caption{Left: electron temperature profile from IMPACT (black line) and CTC using both $f=0.05$ and $f=0.15$ (blue and red lines respectively) after 138 ps laser heating. Right: corresponding electric fields (in the $x$-direction) from IMPACT (black line) and CTC using $f=0.05$ (blue line) and $f=0.15$ (red line).}
\label{temp_efield_VFP}
\end{figure}

Comparison to kinetic IMPACT simulation results, shown in figure \ref{temp_efield_VFP} demonstrates clearly that the electric field in the MHD calculation is dominated by this numerical artefact due to the artificially steepened temperature profile for both $f=0.05$ and $f=0.15$.  Due to the computationally intensive nature of the IMPACT simulations compared to CTC, it was only possible to run these simulations for 138 ps of laser heating.  Nevertheless they demonstrate that the steepening of the profile, seen particularly for the $f=0.05$ CTC simulation is an artefact of the flux limiter.  This has a very large effect on the electric field, which is much reduced and has a completely different profile, i.e. it is not strongly peaked where the flux limiter (artificially) steepens the electron temperature profile.  The electric field from CTC displays what appear to be oscillations in the $f=0.05$ case immediately behind the transport barrier (giving an inversion in the direction of the electric field there).  We believe this is due to the steepness of the temperature profile there and as such is unphysical.

\begin{figure}
\centering\includegraphics[width=2.5in]{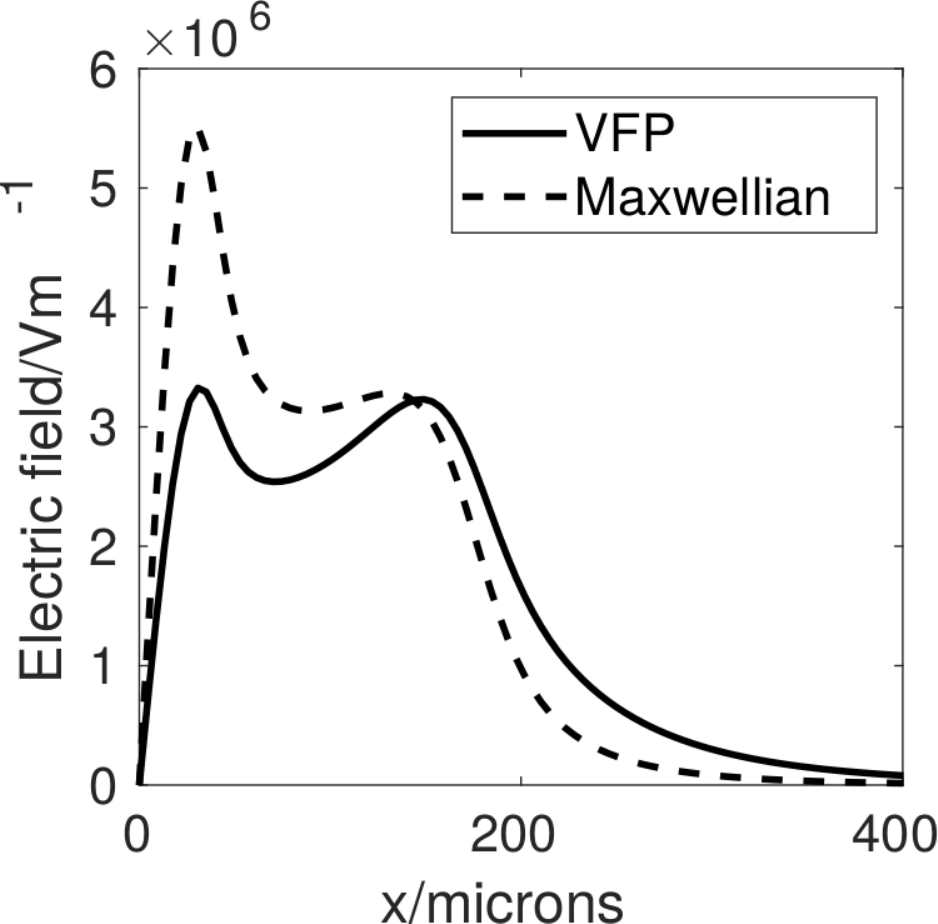}
\caption{Left: electron temperature profile from IMPACT after 138 ps laser heating. Right: corresponding electric field (in the $x$-direction, solid line) \& expected electric field from this temperature profile assuming the electron distribution function is Maxwellian (dashed line).}
\label{Efield_VFP_Max}
\end{figure}

Shown in figure \ref{Efield_VFP_Max} is the electric field produced by assuming the electron distribution function is Maxwellian, i.e. using equation (\ref{local_E}), but taking the electron temperature profile from IMPACT. The difference between this and the electric field from IMPACT shows the direct effect of the distortion of the distribution function away from Maxwellian on the Biermann battery.  While this effect does make a difference to the electric field it is substantially smaller than the effect of the thermal flux limiter.

\subsection{Synthetic proton radiography}

To determine whether kinetic effects on the electric field could be measured in an experiment we have performed synthetic proton radiography \cite{Willingale_10} on the electric fields similar to those produced by CTC and IMPACT shown in figures \ref{temp_efield_CTC} and \ref{Efield_VFP_Max}.  Assuming that such an experiment would have cylindrical symmetry about the $z$-axis we assume the electric field would be radial in the $x,y$ plane.  We have set the magnitude of this radial electric field equal to the electric field from the 1D simulations shown in figures \ref{temp_efield_CTC} and \ref{Efield_VFP_Max}.  Approximate cylindrical symmetry could be achieved experimentally by using a long focal length optic. Cylindrical symmetry ensures that the Biermann-producing electric field is curl-free (assuming the gas density is initially uniform) and therefore no magnetic field is generated to complicate the proton radiography.

In the limit where the electric potential is much smaller than the proton energy, the deflection of the protons is small, and the predicted proton radiograph can be calculated as described by Kugland \emph{et al.}\cite{Kugland_12,Palmer_19}. This is the case here as the electric potential energy is $\sim k_B T_e$ -- as can be seen from equation (\ref{local_E}) -- which figure \ref{temp_efield_CTC} shows is hundreds of eV, whereas protons used in proton radiography have energies $>$ MeV.  In this case the fields are integrated over the direction of the proton trajectories to give a two-dimensional map of proton deflections. These are projected onto a `screen' to find a map of final proton positions against initial position. This mapping defines a change in area described by the determinant of the two-dimensional Jacobian, and hence the change in intensity can be calculated as the reciprocal of this determinant:
\begin{align}
 \frac{I}{I_0} &= \left | \bm{J} \right |^{-1} \equiv \left | \frac{\partial \bm{x'}}{\partial \bm{x}} \right|^{-1}, \nonumber
\end{align}
where positions in the electric field are described by $\bm{x}$ and on the screen by $\bm{x'}$ and:
\begin{align}
 \bm{x'} \approx \bm{x} + L_2 \left(\frac{\bm{x}}{L_1} + \bm{\theta} \right ), \nonumber
\end{align}
for a distance $L_1$ from the proton source to the electric field object, and a distance $L_2$ from the electric fields to the screen. In the case considered here the protons propagate in the $x$-direction with a symmetrical spread of velocities in the $y$-direction and thus are deflected solely in the $y$-direction.  The deflection $\bm{\theta}=(0,\theta_y,0)$ is described by: 
\begin{equation}
    \theta_y \approx \frac{\delta v_y}{v_0} \approx \frac{e}{m_pv_0^2} \int
      E_y \mathrm{d}l. \nonumber
\end{equation}

\begin{figure}
\centering\includegraphics[width=5in]{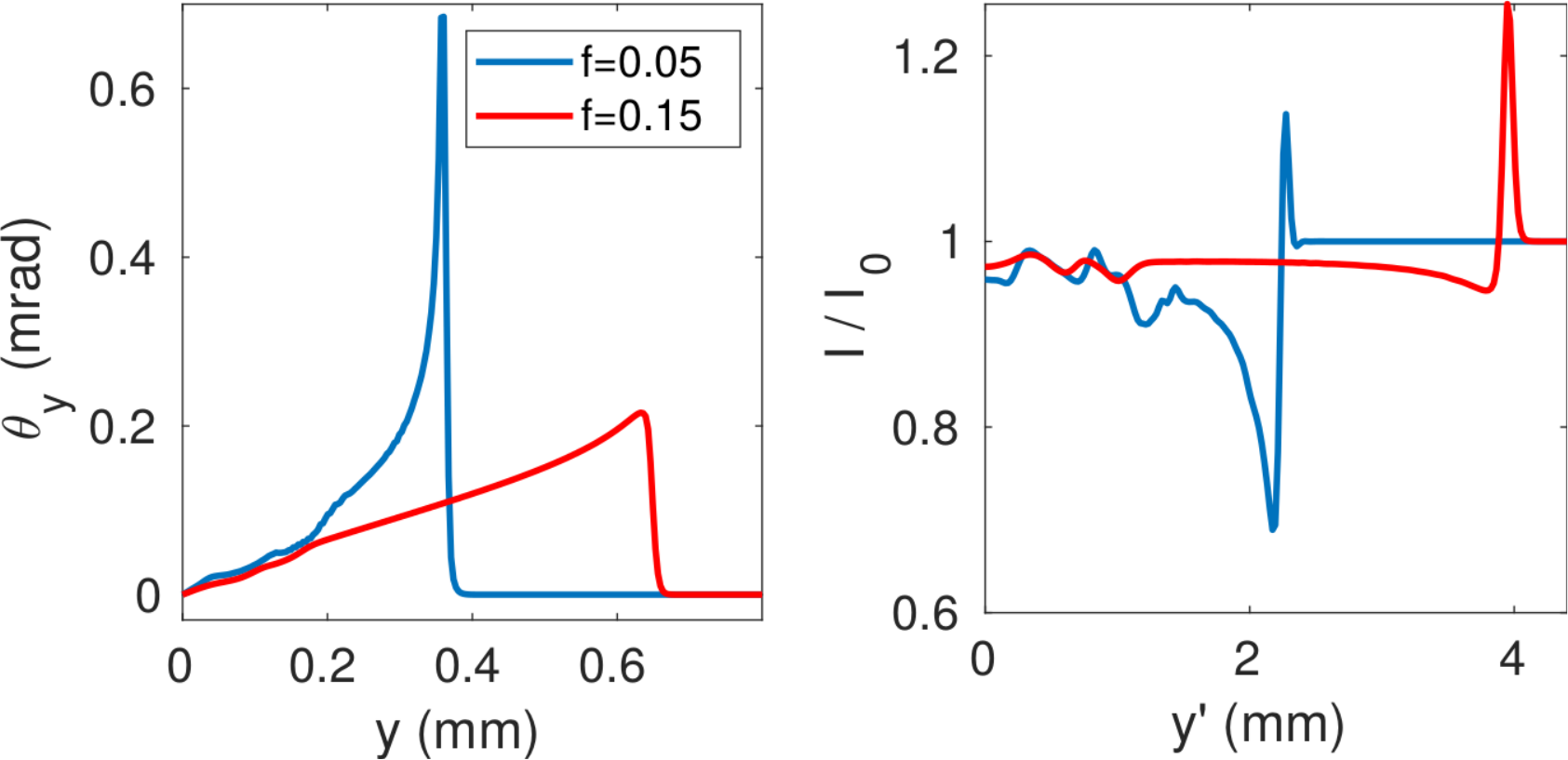}
\caption{Proton deflection (left) and intensity at the screen (right) for CTC simulations after 1 ns of laser heating with $f=0.05$ (blue line) and $f=0.15$ (red line).}
\label{radiographs}
\end{figure}

\begin{figure}
\centering\includegraphics[width=5in]{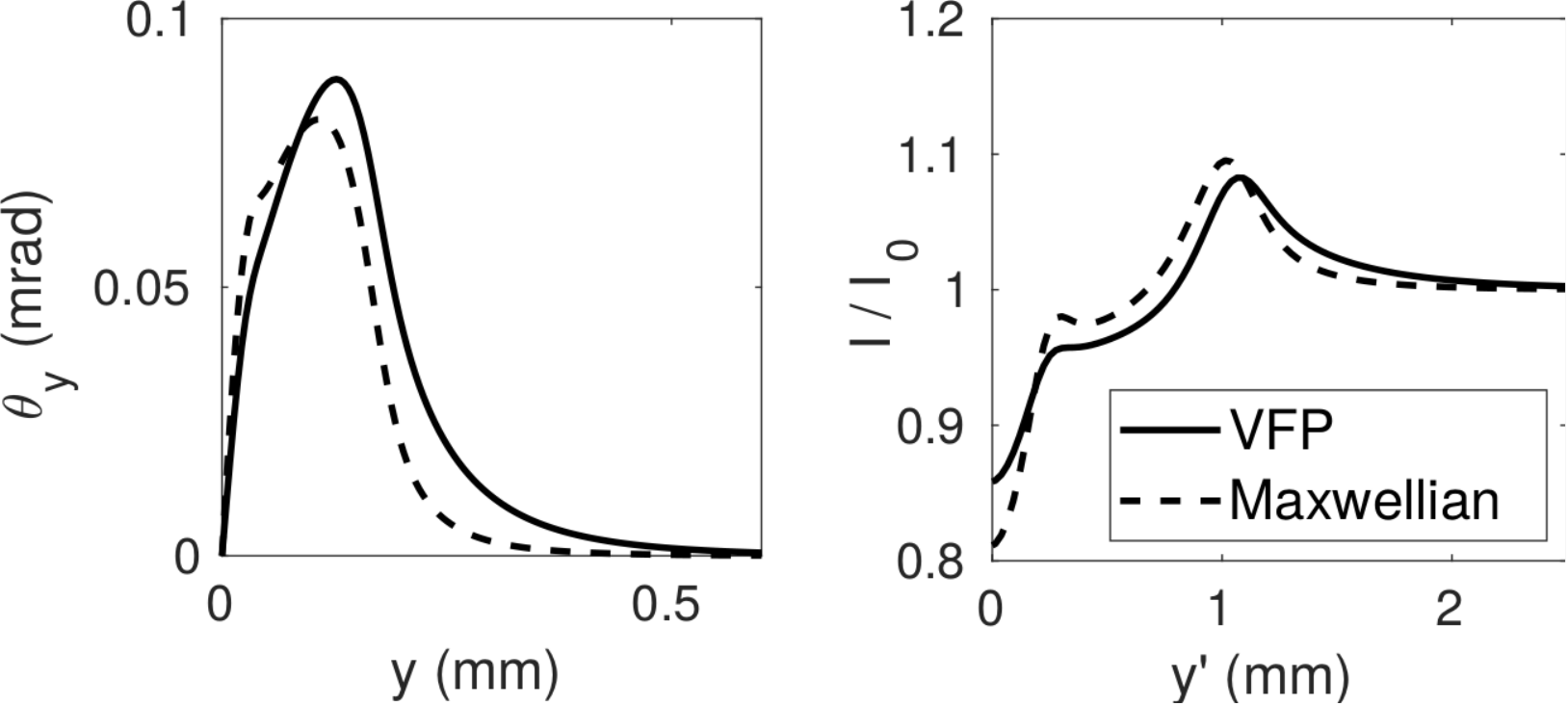}
\caption{Proton deflection (left) and intensity at the screen (right) for IMPACT simulations after 138 ps of laser heating (solid line) and for the electric field assuming the same plasma temperature and density as the IMPACT simulation but assuming the electron distribution function is Maxwellian (dashed line).}
\label{r}
\label{radiographs_VFP}
\end{figure}

\noindent Here $\delta v_y$ is the change in the velocity of the proton (mass $m_p$) in the $y$-direction (initially moving at speed $v_0$) due to the $y$-component of the electric field ($E_y$). We integrate over the electrons trajectory, which is approximately a straight line as the deflections are small. We have assumed the protons are non-relativistic.

The resulting synthetic radiographs for the fields predicted by MHD (based on the 1D CTC simulations shown in figure \ref{temp_efield_CTC}) can be seen in figure \ref{radiographs}.  The synthetic radiographs for fields predicted by kinetic theory (based on the 1D IMPACT simulations shown in figure \ref{Efield_VFP_Max}) are shown in figure \ref{radiographs_VFP}.  We have assumed monoenergetic protons with kinetic energy 5 MeV.  The proton source was assumed to be 100 mm from the plasma; the screen was 500 mm from the plasma.  The proton source was assumed to have a finite size.  We chose to represent this by a Gaussian profile with FWHM $S=11.8$ microns (although this was somewhat arbitrary). To include the effect of this finite source size on the synthetic radiograph we convolve the radiograph with a Gaussian of FWHM $(1+L_2/L_1)S = 70.6$ microns.

A simple estimate demonstrates that the deflection is dominated by the spurious spike in the electic field caused by the flux limiter \footnote{rather than caustics.}. Assuming that this radial electic field is structured as a cylindrically-symmetric shell of radius $r$ and thickness $\Delta L$ then the maximum deflection of the protons is given by $\theta_y\sim (5/2) (\Delta T_e / \mbox{K.E.})\sqrt{r/\Delta L}$ where $\Delta T_e$ is the drop in electron temperature over the artificial (flux limiter induced) transport barrier (thickness $\Delta L$), K.E. is the kinetic energy of the protons. To obtain this we have used equation (\ref{local_E}) and assumed the deflection is small.  Using this formula we obtain $\theta_y\sim 0.6$ mrad for the $f=0.05$ case (where we saw above that CTC gives $\Delta T_e \approx 150$ eV, $r \approx 400$ microns and $\Delta L\approx 6.8$ microns).  For the $f=0.15$ case we obtain $\theta_y \approx 0.1$ mrad (CTC gives $\Delta T_e \approx 50$ eV, $r \approx 600$ microns and $\Delta L\approx 17$ microns).  That these agree with the observed deflections in figure \ref{radiographs} supports the conclusion that the electric field resulting from the flux limiter is responsible for the features in the synthetic radiographs.  This suggests a relatively simple experiment for observing kinetic effects on the Biermann battery.

This conclusion is further supported by the synthetic radiographs based on the IMPACT simulations, shown in figure \ref{radiographs_VFP}.  These display a completely different structure to the synthetic radiographs based on the CTC simulations, notably the spikes are absent, caused by the spikes in the electric field are absent.  The measurement threshold for the intensity changes on the radiographs is estimated at approximately 10\%, thus the deflection of the protons due to the Biermann-producing electic field in the more physical kinetic simulations should be observable.

\subsection{Magnetic field generation in 2D simulations}

To demonstrate the importance of the thermal flux limiter on the magnetic field generated by the Biermann battery we have conducted 2D CTC simulations, with identical physical conditions to the 1D simulations except that a cosine density perturbation was added in the $y$-direction, i.e. the electron density was $n_e = 1.5\times 10^{19}$\ cm$^{-3} [1 - 10^{-3}\cos(2\pi y /200\ \mu \mbox{m}) ] $.  The laser heating profile remained unchanged from the 1D simulations, $I=I_0 e^{-x^2/w^2}$ where $w=20$ microns (so the FWHM  is 33.3 microns).   In this case the electric field generated by the electron pressure gradient will have a curl and a magnetic field will be generated according to Faraday's law $\partial\mathbf{B}/\partial{t} = -\nabla\times\mathbf{E} = -(1/en_e)\nabla_{\mathbf{x}}n_e \times \nabla_{\mathbf{x}} T_e$. A numerical grid 30 $\times$ 30 cells was used to represent 200 microns in each direction.  30,000 time steps were used to discretise a total simulation time of 60\ ps.  In order to prevent numerical instabilities the thermoelectric term had to be artificially switched off. Equation (\ref{local_E}) shows that this will not affect the B-field generation rate as this term is curl-free, although it will curtail B-field transport by the Nernst effect, which would be important in this setup \cite{Ridgers_08}.

\begin{figure}
\centering\includegraphics[width=5in]{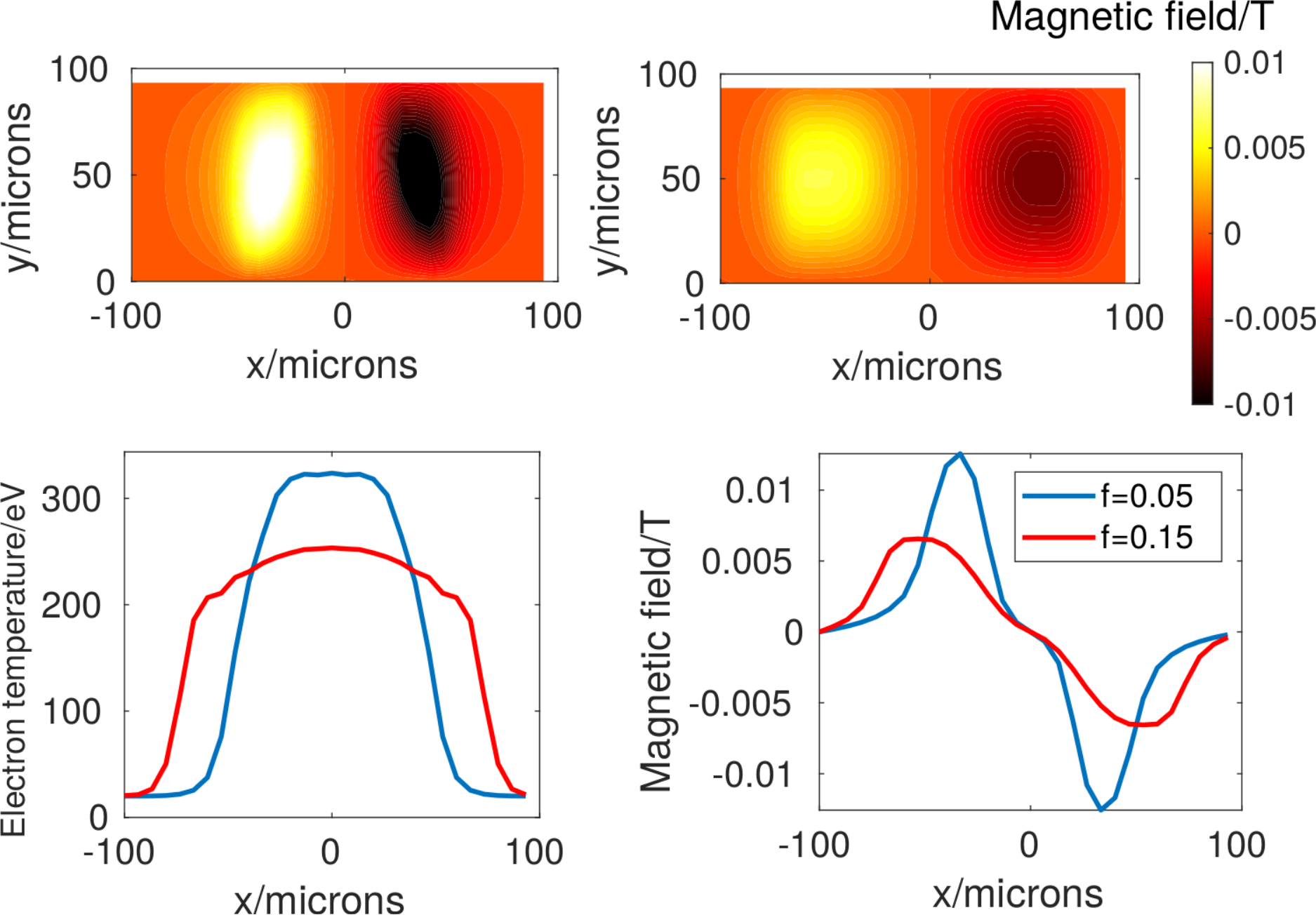}
\caption{Magnetic field from CTC after 60 ps using $f=0.05$ (top left) and $f=0.15$ (top right).  Corresponding electron temperature (bottom right) and lineout of the magnetic field at $y=50$ microns (bottom left).}
\label{2D_B_field}
\end{figure}

Figure \ref{2D_B_field} shows the Biermann-generated magnetic field after 60\ ps using $f=0.05$ and $f=0.15$. Although the resolution of these simulations was relatively poor as the flux limiter caused the $f=0.05$ simulations to go unstable, a clear reduction is seen in B-field generation when the flux limiter is increased from $f=0.05$ and the B-field is more localised. Also shown in figure \ref{2D_B_field} are the electron temperature profiles from the $f=0.05$ and $f=0.15$ CTC simulations after 60 ps as well as lineouts of the B-field through the point of maximum field (at $y = 50$ microns).  The temperature profiles show that the magnetic field is generated at the steep fall in the temperature due to the artificial transport barrier for both $f=0.05$ and $f=0.15$ and is therefore unphysical.  The line out of the B-field shows that the line integral of the magnetic field is reduced by the flux limiter (often this is the important variable), in fact the line integral of the magnetic field from $x=0$ to 100 microns is reduced by 20\% for $f=0.15$ compared to $f=0.05$.

Finally we note that a situation similar to this 2D simulation setup may be realisable experimentally by inducing a density jump in a gas jet.  This could be achieved by overlapping two gas jets of differing pressure or by the presence of an obstacle in the gas jet (although the latter would result in a shock).

\section{Discussion}

Here we have shown that two kinetic effects on the Biermann battery become important when the electron mean free path can no longer be considered small compared to the electron temperature gradient.  This is due to the fact that LTE breaks down in this case and the electron distribution function is no longer Maxwellian as the transport becomes nonlocal.  Firstly there is the direct effect of the distortion of the distribution function on the electric field, which has been previously considered \cite{Kingham_02,Sherlock_20}.  The electric field causing the Biermann battery is given by equation (\ref{nonlocal_E}) in the general case where the distribution function is not Maxwellian.  Only in the special case of LTE where the distribution is Maxwellian is the electric field given by the more familiar form in equation (\ref{local_E}) usually employed in MHD codes.  In the simulations presented in section \ref{simulations}, relevant to the conditions in inertial confinement fusion plasmas, this direct effect of the distortion of the electron distribution reduced the peak electric field by almost 50\%, in line with previous results \cite{Kingham_02}.

The second, less explored kinetic effect on the Biermann battery is the indirect effect of the flux limiter on electron heat transport, also commonly employed in MHD codes.  We showed that varying the electron flux limiter between reasonable values (in the context of inertial confinement fusion) lead to more than a factor of six change in the instantaneous peak electric field in 1D simulations after 1 ns of heating of a nitrogen gas jet.  We expect this difference would grow on longer timescales.  Moreover the electric field profile was dominated by a very narrow peak in this case (see figure \ref{temp_efield_CTC}) which is a numerical artefact of the flux limiter.  This is due to the fact that the flux limiter causes an artificial transport barrier, steepening the temperature profile unphysically, leading to the strong peak in the electric field.  Figure \ref{temp_efield_VFP} shows that this feature is absent from the more physically correct VFP simulations.  2D CTC simulations showed that this unphysical transport barrier did indeed modify the rate of B-field generation by the Biermann battery, leading to an increase in the magnitude of the magnetic field by a factor of two after 60 ps of laser heating for $f=0.05$ compared to $f=0.15$.  This suggests that our MHD predictions of the magnetic field from the Biermann battery are dominated by a numerical artefact (the flux limiter induced transport barrier).  In addition in the $f=0.05$ case the electron temperature profile steepened so much that the temperature dropped by a large fraction of the peak over a few grid cells.  This would be expected to introduce not only inaccuracies in calculating the electric field, but the thermal transport as well.

While the direct effect of nonlocality on the Biermann battery is interesting the indirect effect of the flux limiter is more important. There is no single value for the thermal flux limiter $f$ which accurately captures kinetic effects.  The fact then that MHD simulations of the Biermann battery should depend so strongly on $f$ therefore suggests that care should be exercised interpreting the results of such simulations.  For example, MHD simulations of indirect drive ICF suggest that large magnetic fields may be generated by the Biermann battery but the effect of the flux limiter ($f=0.15$ was used) on these predictions was not explored \cite{Farmer_17}.  Recently particle in cell kinetic simulations have been used to demonstrate further departures from simple MHD modelling, for example the formation of filaments due to the Weibel instability \cite{Schoeffler_18,Shukla_20}.  This effect is not included in our modelling as it is reliant on higher order anisotropy in the electron distribution function.  Indeed the IMPACT simulation results present here are at the limits of the validity of truncation of the electron distribution at first order anisotropy.  Higher order terms are negligible if the ratio of the scale length to the mean free path is small \cite{Johnston_60}, yet as an example they are of approximately the same magnitude at the heat front (defined as being at the half maximum of the temperature profile) for the IMPACT-produced temperature profile marked `VFP' in figure \ref{temp_efield_VFP}.  While this may affect the details of the thermal transport (interesting as a subject for further work) we do not expect it to change the key qualitative result that flux limited MHD perdicts an artificaially steep temperature profile. 

We have also not discussed the distortion of the electron distribution function due to inverse bremsstrahlung heating (which pushes the distribution towards a super-gaussian).  This can modify transport \cite{Ridgers_08_2}, suppressing the Biermann battery by up to 30\% \cite{Bissell_13}).   This is not expected to be important here as the Langdon parameter $Zv_{\mbox{osc}}^2/v_T^2\approx 0.4$ where $v_{\mbox{osc}}$ is the quiver velocity of the electrons in the laser field and we have assumed a temperature of 300 eV (from figure \ref{temp_efield_VFP}).

Our simulations have other limitations.  We have limited our kinetic simulations to one spatial dimension.  Two dimensional kinetic simulations would not only be computationally intensive but in order to meaningfully compare to the MHD simulations the Nernst effect would need to be artificially removed, which is not straightforward in a kinetic framework.  While the 1D assumption artificially imposes symmetry not present in the experiment proposed in section \ref{simulations}(c) we would expect qualitatively similar results in these situations as the proposed experiment has cylindrical symmetry meaning the Biermann-producing electric field will remain curl-free and no magnetic field will be generated.  This is in contrast to the experiment proposed in section \ref{simulations}(d), where this symmetry is broken and B-fields are generated, although in the simuations Cartesian rather than cylindrical coordinates were again used.  We have also neglected discussing the subsequent transport of the magnetic fields. This occurs primarily by the Nernst effect, whereby the magnetic field is advected by the electron heat flow \cite{Willingale_10}. As the electron heat flow must be described kinetically, then so must the Nernst effect and therefore the magnetic field advection \cite{Ridgers_08,Joglekar_16}.  In addition effects such as the Nernst effect can give rise to novel plasma instabilities \cite{Bissell_10} making this experimental platform potentially very fruitful for studying kinetic effects on transport (but making modelling challenging -- we had to artificially switch the Nernst term off to get the 2D CTC simulations to remain numerically stable).

The significant kinetic effects on the Biermann battery provide motivation to validate our modelling with experiment.  We have proposed a simple experiment where a ns pulse length laser of intensity $\sim 10^{14}$ Wcm$^{-2}$ heats an underdense gas (electron density $\sim 10^{19}$ cm$^{-3}$) similar to previous experiments \cite{Gregori_04,Froula_07} (and suggested experiments \cite{Walsh_20}). Synthetic proton radiography shows that such an experiment can demonstrate the inaccuracy of MHD modelling of the Biermann battery when a thermal flux limiter is used, although the direct effect of the distortion of the electron distribution function is smaller and so harder to observe, requiring a proton radiography setup at the limit of size that could be fielded experimentally for sufficient magnification (though measurement of the indirect effect would not require such a high magnification).  By inducing a density variation in the gas jet the Biermann-producing electric field will have a curl and a magentic field will be generated. This could provide a route to experimentally observing the direct and indirect kinetic effects discussed here on the Biermann-generated magnetic field itself rather than just the electric field which produces it as the former is usually of more interest.

Determining whether kinetic effects can be incorporated into the MHD framework to sufficient accuracy is very important as full kinetic codes are currently too computationally intensive for full scale simulations of, for example, an ICF experiment. Recent work suggests that reduced kinetic models of electron thermal transport provide a compromise which is sufficiently accurate and efficient \cite{Marrochino_13,Brodrick_17,Sherlock_17}.  While these models have been shown to be suitable for describing the Nernst effect \cite{Brodrick_18} further work is required to determine whether they can describe the Biermann battery.

\section{Conclusions}
We have shown that kinetic effects modify the Biermann battery under conditions relevant to current laser plasma experiments (and inertial confinement fusion).  While the direct effect on this electric field from the distortion of the electron distribution function away from Maxwellian was found to lead to a decrease in the peak electric field by approximately 50\%, the dominant effect is the artificial steepening of the temperature profile by the flux limiter.  This latter effect meant that magnetothydrodynamics, often used to model the Biermann battery, produced fields dominated by a numerical artefact from this steepening.  We have shown that this inadequacy of MHD to correctly model the Biermann battery is observable experimentally, suggesting a strategy for much needed benchmarking which can be performed using current high power (ns pulse) laser systems.  

\enlargethispage{20pt}

Data managed by University of York library via doi 10.15124/f81d0a18-19c8-4ae5-890e-db8c55747779

CPR performed the IMPACT and CTC simulations with support from RJK and JJB respectively.  CA performed the synthetic proton radiography.

The authors declare that they have no competing interests.

CPR would like to acknowledge funding from the UK Engineering and Physical Sciences Research Council (grant EP/M011372/1).  This work has been carried out within the framework of the EUROfusion Consortium and has received funding from the Euratom research and training programme 2014–2018 under Grant Agreement No.633053 (Project Reference CfP-AWP17-IFE-CCFE-01). The views and opinions expressed herein do not necessarily reflect those of the European Commission.


\end{document}